\documentclass[preprint,superscriptaddress,amsmath,amssymb,prd,aps,showpacs,floatfix,nofootinbib]{revtex4-1}

\usepackage[T1]{fontenc}
\usepackage{lmodern}

\usepackage[dvipsnames,x11names]{xcolor}

\usepackage[colorlinks=true,linkcolor=Blue,citecolor=ForestGreen,urlcolor=NavyBlue]{hyperref}

\usepackage{graphicx}
\usepackage[sort&compress]{natbib}
\usepackage{lipsum}
\usepackage{morefloats}
\usepackage[pdf]{pstricks}
\usepackage{subfigure}
\usepackage{amsmath}
\usepackage{amssymb}
\usepackage{amsfonts}
\usepackage{xcolor}
\usepackage{rotating}
\usepackage{cancel}
\usepackage{mathtools}
\usepackage{bbm}
\usepackage{dsfont}
\usepackage{bbold}
\usepackage{multirow}
\usepackage{ulem}
\usepackage{physics}
\usepackage{orcidlink}

\newcommand{\mev}{\textrm{ MeV}}

\newcommand{\GXNU}{Department of Physics, Guangxi Normal University, Guilin 541004, China}
\newcommand{\GXZD}{Guangxi Key Laboratory of Nuclear Physics and Technology, Guangxi Normal University, Guilin 541004, China}
\newcommand{\IFIC}{Departamento de F\'{\i}sica Te\'orica and IFIC, Centro Mixto Universidad de
Valencia-CSIC Institutos de Investigaci\'on de Paterna, Aptdo.22085,
46071 Valencia, Spain}

\begin{document}

\frenchspacing

\title{Correlation function and the inverse problem in the $BD$ interaction}

\author{Hai-Peng Li}
\affiliation{\GXNU}%

\author{Jing-Yu Yi}
\affiliation{School of Physics and Electronics, Hunan University, 410082 Changsha, China}%

\author{Chu-Wen Xiao}
\email{xiaochw@gxnu.edu.cn}
\affiliation{\GXNU}%
\affiliation{\GXZD}%
\affiliation{School of Physics, Central South University, Changsha 410083, China}%

\author{De-Liang~Yao}
\email{yaodeliang@hnu.edu.cn}
\affiliation{School of Physics and Electronics, Hunan University, 410082 Changsha, China}%

\author{Wei-Hong Liang}%
\email{liangwh@gxnu.edu.cn}
\affiliation{\GXNU}%
\affiliation{\GXZD}%

\author{Eulogio Oset}%
\email{Oset@ific.uv.es}
\affiliation{\GXNU}%
\affiliation{\IFIC}%

\begin{abstract}
  We study the correlation functions of the $B^0 D^+, B^+ D^0$ system, which develops a bound state of approximately $40\mev$, using inputs consistent with the $T_{cc}(3875)$ state. Then we address the inverse problem starting from these correlation functions to determine the scattering observables related to the system, including the existence of the bound state and its molecular nature. The important output of the approach is the uncertainty with which these observables can be obtained, considering errors in the $B^0 D^+, B^+ D^0$ correlation functions typical of current values in present correlation functions. We find that it is possible to obtain scattering lengths and effective ranges with relative high precision and the existence of a bound state. Although the pole position is obtained with errors of the order of $50 \%$ of the binding energy, the molecular probability of the state is obtained with a very small error of the order of $6\%$. All these findings can serve as motivation to perform such measurements in future runs of high energy hadron collisions. 
\end{abstract}

\maketitle

\section{Introduction}
\label{sec:intro}

Femtoscopic correlation functions are emerging as a tool for understanding the interaction of hadrons at small relative momenta.
Considerable experimental research has already been conducted within the strange sector \cite{stara,alice1,alice2,starb,alice3,alice4,alice6,alice5,alice7,alice9,alice8,expe1,expe2,ALICE:2022yyh} (see also review paper \cite{fab}).
Incursions in the $D$ sector have been investigated \cite{alice10} and plans to extend studies to this sector are expected in the near future \cite{ALICE:2022wwr}.
In future runs of the LHC, the ALICE collaboration will likely also access the bottom sector.

The theoretical community is also devoting considerable research to the subject \cite{morita,ohnishi,feijoc,morita1,hatsuda,mihaylov,haiden,morita2,kamiya,kamiya1,kamiya2,liuwei,vidafe,albal,Liulu,liuwen}, and a model independent analysis of the correlation functions was very recently proposed \cite{ikeno,scibull,Feijootbb,Molinanstar,Molina:2023oeu}, where instead of contrasting theoretical models with experimental data, the inverse path was followed. The data were used to determine the scattering observables of coupled channels, explore the possibility of having several bound states and eventually determine the nature of these bound states as molecular states or otherwise.

In the present study, we address the $BD$ interaction, where a bound state was predicted for the $BD$ system with isospin $I=0$ in Ref.~\cite{Sakai:2017avl}. The observation of the $T_{cc}$ state \cite{LHCb:2021vvq,LHCb:2021auc} and subsequent theoretical research support this state as a bound state of the $D^{0}D^{*+}$, $D^{+}D^{*0}$ channels in $I=0$ \cite{Li:2012ss,Liu:2019stu,Ding:2020dio,Feijoo:2021ppq,Dong:2021bvy,Chen:2021cfl,Albaladejo:2021vln,Du:2021zzh,Agaev:2022ast,Chen:2022asf,Wang:2023ovj,Chen:2023fgl}.
Two recent studies investigating $T_{cc}$ from a different perspective also concluded that the $T_{cc}$ state is indeed a molecular state \cite{Dai:2023cyo,Dai:2023kwv}.
In the first study, the scattering length and effective range of the $D^{0}D^{*+}$, $D^{+}D^{*0}$ channels, as well as the shape of the $D^{0}D^{0}\pi^{+}$ mass distributions were theoretically studied \cite{Dai:2023cyo}, concluding that $T_{cc}$ is a molecular state of the $D^{0}D^{*+}$, $D^{+}D^{*0}$ components, essentially constructing an $I=0$ state.
In the second study, a different path was followed, assuming that $T_{cc}$ could correspond to a non molecular, genuine or preexisting state and be dressed by the meson-meson components where it is observed \cite{Dai:2023kwv}. It was found that, in principle, it would be possible to have a $T_{cc}$ state with negligible molecular probability, but at the heavy cost of very small $D^{0}D^{*+}$ scattering length and huge effective range, which are far from the observed experimental values.

If $T_{cc}$ is bound, using arguments of heavy quark symmetry, the $\bar{B}D^{*}$ state should also be bound, even more than $T_{cc}$ because there is a general rule that the heavier the quarks, the stronger the interaction and binding energies \cite{Ader:1981db,Zouzou:1986qh,Carlson:1987hh}.
This rule is also satisfied when the extension of the local hidden gauge approach \cite{Bando:1984ej,Furui:1995bj,Meissner:1987ge,Nagahiro:2008cv} is used as a source of interaction, exchanging vector mesons between the heavy meson components. In Ref.~\cite{Sakai:2017avl}, different $B^{(*)}D^{(*)}$, $B^{*}\bar{D}^{(*)}$ pairs were found to be bound using the same regulator for the loops as in Ref.~\cite{Feijoo:2021ppq}, particularly the $BD$ state in $I=0$, which was found to be bound by approximately $15-30 \mev$.
The existence of one bound state in the $BD$ system is also supported by the phase moment obained in Ref.~\cite{Yao:2019vty}.

The finding of the $T_{cc}$ state has been extremely useful for placing constraints on the meson-meson interaction and its range, as reflected in the cutoff used to regularize the loops \cite{Gamermann2010,Song:2022yvz}.
The information and arguments on heavy quark symmetry used in Ref.~\cite{Sakai:2017avl} consolidated their results.
With confidence in the predictions of Ref.~\cite{Sakai:2017avl} and their inputs consistent with the information obtained from $T_{cc}$, we present here one work which should stimulate experimental measurements to corroborate these findings.
First, we reproduce the result of Ref.~\cite{Sakai:2017avl} and evaluate the correlation functions of $D^{0}B^{+}$ and $D^{+}B^{0}$.
Next, assuming that these correlation functions correspond to actual data, we address the inverse problem of determining from them the value of the scattering observables, scattering length and effective range for these two channels; the existence of a bound state below the threshold; the molecular probability of each of the $D^{0}B^{+}$ and $D^{+}B^{0}$ components; and most importantly, the precision with which we can determine these magnitudes, assuming reasonable error bars for the correlation function data. This information is important to obtain an idea of what to expect given the experimental constraints when such experiments are performed.

\section{Formalism}
\label{sec:forma}
\subsection{Interaction of the $D^{0}B^{+}$, $D^{+}B^{0}$ channels}
\begin{figure}[t]
     \centering
    \includegraphics[width=1.05\linewidth]{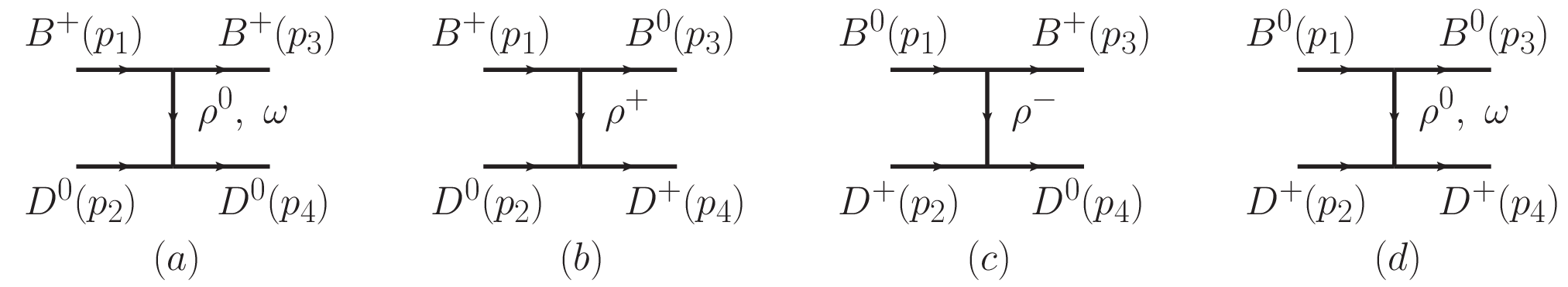}
    \caption{Vector exchange between the $D^{0}B^{+}$ and $D^{+}B^{0}$ components (in brackets, the four momentum of each particle).}
 \label{fig:Feyn}
\end{figure}

The details can be found in Ref.~\cite{Sakai:2017avl};
hence, we do not cover them here. Light vector meson are exchanged between the $B$ and $D$ components, as shown in Fig.~\ref{fig:Feyn}.
We can see that the heavy quarks are spectators in the exchange of light vector mesons.
Furthermore, assuming $m_{\rho}=m_{\omega}$, the interaction for the $B^{+}D^{0}(1)$, $B^{0}D^{+}(2)$ channels is given by
\begin{equation}\label{eq:Vij}
	V_{ij}=-\frac{1}{4f^2}\;C_{ij}\;(p_{1}+p_{3}) \cdot (p_{2}+p_{4});\quad f=93\ \text{MeV},
\end{equation}
with $C_{ij}$ the element of matrix $C$,
\begin{equation}\label{eq:Cij}
	C=
	\begin{pmatrix}
		1 & 1 \\
		1 & 1 \\
	\end{pmatrix},
\end{equation}
and projected over $S$-wave
\begin{equation}\label{eq:pij}
	(p_{1}+p_{3}) \cdot (p_{2}+p_{4})\to\frac{1}{2}\left[3s-2(m^2_{B}+m^2_{D})-\frac{(m_{B}^2-m_{D}^2)^2}{s}\right].
\end{equation}
The scattering matrix is then given by
\begin{equation}\label{eq:BS}
	T=[1-VG]^{-1}V,
\end{equation}
where $G={\rm diag}(G_{1},G_{2})$, with $G_{i}$ being the $BD$ meson loop function regularized with a cutoff
\begin{equation}\label{eq:loop}
  G_i(\sqrt{s})= \int_{|\vec q\,| < q_{\rm max}} \dfrac{{\rm d}^3 q}{(2\pi)^3} \; \dfrac{\omega_1(q)+\omega_2(q)}{2\,\omega_1(q)\, \omega_2(q)}\; \dfrac{1}{s-[\omega_1(q)+\omega_2(q)]^2 + i \varepsilon},
\end{equation}
where $\omega_1(q)=\sqrt{\vec q^{\;2} +m_i^2}$, $\omega_2(q)=\sqrt{\vec q^{\;2} +M_i^2}$, and $m_i, M_i$ are the masses of the $D$ and $B$ mesons in channel $i$.
With the isospin states $(B^{+}, B^{0})$, $(D^{+}, -D^{0})$, the $I=0$ combination is given by
\begin{equation}\label{eq:BD1}
	\ket{BD, I=0}=-\frac{1}{\sqrt{2}}\ket{B^{+}D^{0}+B^{0}D^{+}},
\end{equation}
with which we can obtain
\begin{equation}\label{eq:BD2}
	\mel{BD, I=0}{V}{BD, I=0}=\frac{1}{2}(V_{11}+V_{22}+2V_{12})=2V_{11},
\end{equation}
which is the result obtained in Ref.~\cite{Sakai:2017avl}, with $V_{ij}$ given by Eqs.~\eqref{eq:Vij}, \eqref{eq:Cij}, \eqref{eq:pij}. The $I=1$ combination is given by
\begin{equation}\label{eq:BD3}
	\ket{BD, I=1,I_{3}=0}=-\frac{1}{\sqrt{2}}\ket{B^{+}D^{0}-B^{0}D^{+}}.
\end{equation}

We reproduce the results of Ref.~\cite{Sakai:2017avl} using the cutoff regularization with an extra form factor (Eq.~(27) from Ref.~\cite{Sakai:2017avl})
stemming from the $S$-wave projection of a vector meson exchange.
However, practically identical results are obtained by ignoring this form factor and decreasing $q_{\text{max}}$.
To match the formalism of the correlation functions of Ref.~\cite{vidafe}, we ignore this form factor and take $q_{\text{max}}$ of the order of $420 \mev$, as used in Ref.~\cite{Feijoo:2021ppq}, to obtain the binding of the related $T_{cc}$ state.

\subsection{Scattering observables}
We can evaluate the scattering length $a$, $r_{0}$ for the $B^{+}D^{0}$ and $B^{0}D^{+}$ channels by recalling the relationship between our $T$ matrix and that used in quantum mechanics \cite{Gamermann2010}
\begin{equation}
	T=-8\pi\sqrt{s}\;f^{QM}\simeq -8\pi\sqrt{s}\;\frac{1}{-\frac{1}{a}+\frac{1}{2}r_{0}k^2-ik},
\end{equation}
with
\begin{equation}
	k=\frac{\lambda^{1/2}(s,m_{1}^2,m_{2}^2)}{2\sqrt{s}},
\end{equation}
from which we easily find
\begin{equation}\label{eq:aj2}
	-\frac{1}{a_{i}}=\left.\left(-8\pi\sqrt{s}\;T_{ii}^{-1}\right)\right|_{s_{{\rm th},i}},
\end{equation}
\begin{equation}\label{eq:r0j}
	r_{0,i}=\left[\frac{2\sqrt{s}}{\mu_{i}}\frac{\partial}{\partial s}\left(-8\pi\sqrt{s}\;T_{ii}^{-1}+ik_{i}\right)\right]_{s_{{\rm th},i}},
\end{equation}
where $\mu_{i}$ is the reduced mass in channel $i$, and $s_{{\rm th},i}$ and $k_{i}$ are the square of the threshold mass and the center of mass momenta of the mesons for channel $i$, respectively.

\subsection{Couplings and probabilities}
We find that there is a pole, which is below the threshold of the two channels and hence corresponds to a bound state.
The couplings are obtained from the $T$ matrix in the vicinity of the pole
\begin{equation}
	T_{ij}=\frac{g_{i}\,g_{j}}{s-s_{p}},
\end{equation}
where $s_p$ is the square of the mass at the pole. Thus
\begin{equation}
	g_{1}^2=\lim_{s\to s_{p}}(s-s_{p})\;T_{11},
\end{equation}
\begin{equation}
	g_{1}\,g_{j}=\lim_{s\to s_{p}}(s-s_{p})\;T_{1j},
\end{equation}
which determine the relative sign of $g_2$ with respect to $g_1$.
Once the couplings are evaluated, we calculate the molecular probabilities of the $B^+ D^0$ and $B^0 D^+$  channels, as in Refs.~\cite{Gamermann2010,Hyodo:2011qc,Sekihara:2014kya}
\begin{equation}\label{eq:Pi}
	\mathcal{P}_{i}=-g_{i}^2\left.\frac{\partial G_{i}}{\partial s}\right|_{s=s_{p}}.
\end{equation}
Another magnitude of relevance is the wave function at the origin in coordinate space given by \cite{Gamermann2010}
\begin{equation}
	\psi_{i}(r=0)=\left.g_{i}\,G_{i}\right|_{s_{{\rm th},i}}.
\end{equation}

\subsection{Correlation functions}

We follow the formalism of Ref.~\cite{vidafe} and write the correlation functions for the two channels as
\begin{eqnarray}\label{eq:C1}
  C_{B^+D^0} (p_{D^0})&=& 1+4\,\pi\, \theta(q_{\rm max}-p_{D^0})\, \int dr \, r^2 S_{12}(r)  \cdot \nonumber\\[2mm]
  && \left\{ \left|j_0(p_{D^0}\, r)+T_{B^+D^0, B^+D^0}(E)\; \tilde{G}^{(B^+D^0)}(r; E)\right|^2 \right.  \nonumber\\[2mm]
  &&\;\left. + \left|T_{B^0D^+, B^+D^0}(E)\; \tilde{G}^{(B^0D^+)}(r; E) \right|^2  - j_0^2 (p_{D^0}\, r) \right\},
 \end{eqnarray}
\begin{eqnarray}\label{eq:C2}
  C_{B^0D^+} (p_{D^+})&=& 1+4\,\pi\, \theta(q_{\rm max}-p_{D^+})\, \int dr \, r^2 S_{12}(r)  \cdot \nonumber\\[2mm]
  && \left\{ \left|j_0(p_{D^+}\, r)+T_{B^0D^+, B^0D^+}(E)\; \tilde{G}^{(B^0D^+)}(r; E)\right|^2 \right.  \nonumber\\[2mm]
  &&\;\left. + \left|T_{B^+D^0, B^0D^+}(E)\; \tilde{G}^{(B^+D^0)}(r; E) \right|^2  - j_0^2 (p_{D^+}\, r) \right\},
 \end{eqnarray}
where $p_i$ is the momentum of the particles in the rest frame of the pair,
\begin{equation}\label{eq:pi}
  p_i=\dfrac{\lambda^{1/2}(s, m_i^2, M_i^2)}{2\, \sqrt{s}},
\end{equation}
$S_{12}(r)$ is the source function,
parameterized as a Gaussian normalized to $1$,
\begin{equation}\label{eq:S12}
  S_{12}(r)= \dfrac{1}{(\sqrt{4\pi}\, R)^3}  \; e^{-(r^2/4R^2)},
\end{equation}
and the $\tilde{G}^{(i)}(r; E)$ function is defined as
\begin{equation}\label{eq:G2}
  \tilde{G}^{(i)}(r; E)= \int \dfrac{{\rm d}^3 q}{(2\pi)^3} \; \dfrac{\omega_1(q)+\omega_2(q)}{2\,\omega_1(q)\, \omega_2(q)}\; \dfrac{j_0(q\, r)}{s-[\omega_1(q)+\omega_2(q)]^2 + i \varepsilon},
\end{equation}
where $j_0(q\, r)$ is the spherical Bessel function, and $E=\sqrt{s} = \sqrt{m_i^2 + \vec p_i^{\; 2}} + \sqrt{M_i^2 + \vec p_i^{\; 2}}$.

\section{Inverse problem: Model independent analysis of the correlation functions}

Here, we assume that the correlation functions have already been measured and attempt to extract the maximum information available using a general framework in which no model assumptions are made.
To perform the test, we use the correlation function with the model described in the previous sections, assuming errors at the order of $\pm 0.02$, which are slightly larger than those obtained in present measurements of correlation functions.

We begin by assuming that there is an interaction between the coupled channels given by
\begin{equation}
	V=
	\begin{pmatrix}
		V_{11} & V_{12} \\
		V_{12} & V_{22} \\
	\end{pmatrix},
\end{equation}
where $V_{ij}$ are unknown potentials to be determined, and the $T$ matrix is given by Eq.~\eqref{eq:BS}, using the $G$ function of Eq.~\eqref{eq:loop} with an unknown $q_{\rm max}$.
We make no assumption on the isospin structure of a possible bound state but assume that the interaction is isospin symmetric, which implies, according to Eqs.~\eqref{eq:BD1},\eqref{eq:BD2}, and\eqref{eq:BD3}, that
\begin{equation}
	\mel{BD, I=0}{V}{BD, I=1}=\frac{1}{2}(V_{11}-V_{22})=0,
\end{equation}
\begin{equation}
	V_{22}=V_{11}.
\end{equation}
In addition, to consider possible sources of interaction originating from channels neglected in our approach, we introduce several energy dependent terms, as discussed in Refs.~\cite{Aceti:2014ala,Hyodo:2013nka} and studied in Refs.~\cite{ikeno,scibull,Feijootbb}:
\begin{equation}
	V_{11}=V_{11}'+\frac{\alpha}{m_{V}^2}(s-s_{{\rm th},1}),
\end{equation}
\begin{equation}
	V_{12}=V_{12}'+\frac{\beta}{m_{V}^2}(s-s_{{\rm th},1}),
\end{equation}
where the factor $m^2_V$ (with $m_V=800\mev$) is introduced to make $\alpha, \beta$ dimensionless.
Then we have $4$ free parameters for the interaction, as well as $q_{\rm max}$ and $R$, which is a total of $6$ parameters to fit the two correlation functions.
One must be aware that there are strong correlations between these parameters because the input used to obtain the correlation functions corresponds to an interaction at $I=0$; hence what matters is the combination $V_{11}+V_{12}$.
This means that the values we obtain for the parameters in fits to the pseudodata are not too meaningful, only the values of the observables obtained from them are significant. 
In order to deal with these correlations we use the bootstrap or resampling method \cite{Press:1992zz,Efron:1986hys,Albaladejo:2016hae}, generating random centroids of the data with a Gaussian distribution weight and performing a large number of fits to the data with the new centroids and same errors.
After each fit, the values of the observables are evaluated and the average and dispersion for each are calculated.

\section{Results}
\label{sec:res}

\begin{figure}[t]
\begin{center}
\includegraphics[scale=0.45]{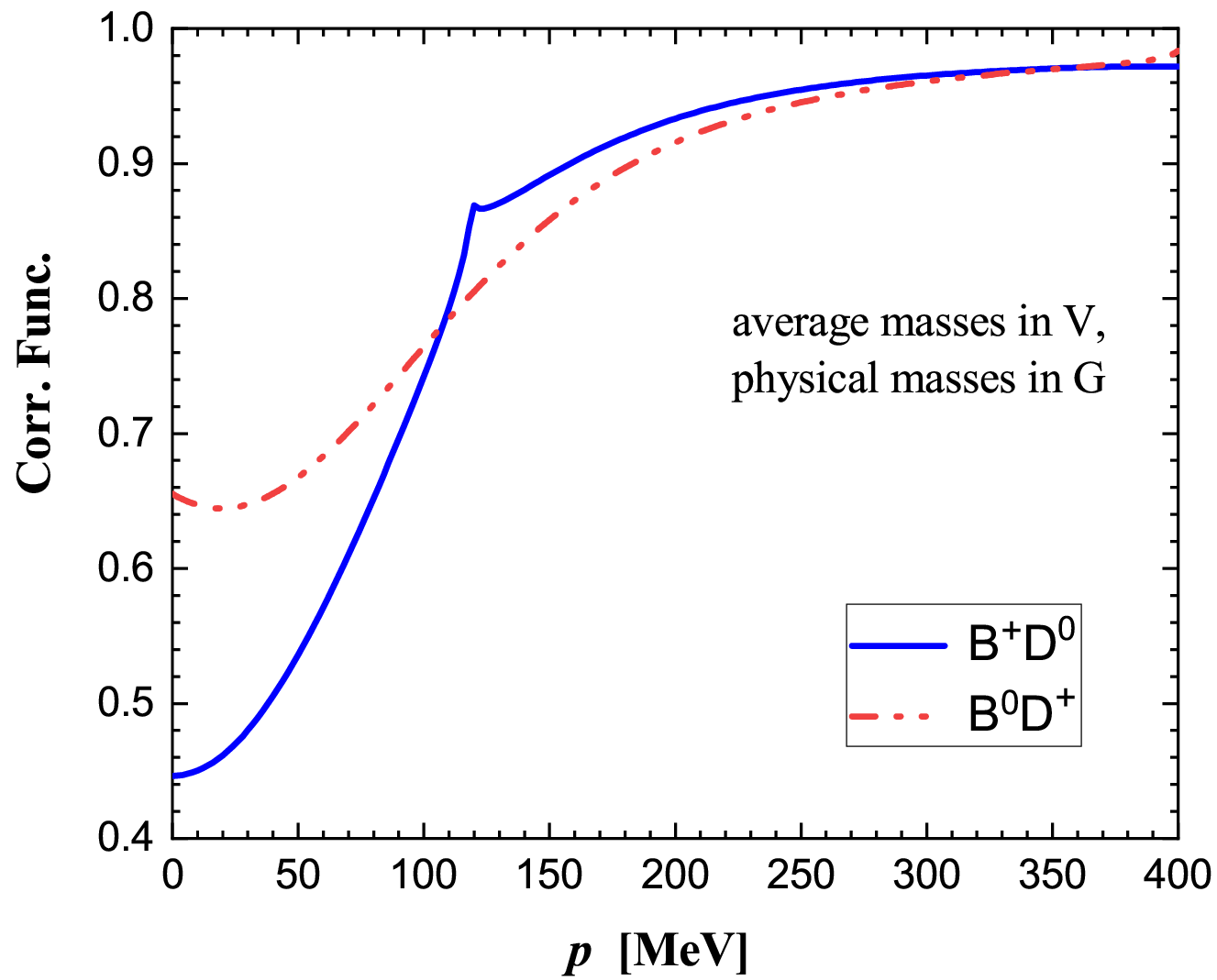}
\end{center}
\vspace{-0.7cm}
\caption{Correlation functions for $B^+D^0$ and $B^0D^+$ channels.}
\label{Fig:CF}
\end{figure}

Next, we use the model in section \ref{sec:forma}, with a cutoff regularization of $q_{\rm max}=420\mev$ as in the study of the $T_{cc}$ state in Ref.~\cite{Feijoo:2021ppq}.
We obtain a pole at $\sqrt{s}=7110.41\mev$ and the couplings given in Table \ref{tab:gi4}. Similarly, the probabilities
 obtained and the wave functions at the origin are given in Table \ref{tab:psii}, and the scattering length and effective range are shown in Table \ref{tab:ai}.
\begin{table}[t]
     \renewcommand{\arraystretch}{0.9}
     \setlength{\tabcolsep}{0.3cm}
\centering
\caption{ Pole position and couplings with $q_{\rm max}=420\; {\rm MeV}$. [in units of MeV]}
\label{tab:gi4}
\begin{tabular}{ccc}
\hline
\hline
$\sqrt{s_p}$ & $g_1$  & $g_2$  \\
\hline
   $(7110.41+0i)$  & $31636.8$ & $31631.0$   \\
\hline\hline
\end{tabular}
\end{table}
\begin{table}[t]
	\caption{Probability $\mathcal{P}_i$ and wave function at the origin $\psi_i(r=0)$ for channel $i$.}
\centering
\begin{tabular*}{0.5\textwidth}{@{\extracolsep{\fill}}cccc }
\toprule
$\mathcal{P}_1$       & ~~$\mathcal{P}_2$~~   & ~~$\psi_1(r=0)$~~     & ~~$\psi_2(r=0)$~~  \\
\hline
~~~$0.52$~~   & $0.44$   & $-14.75$     &  $-13.61$    \\
\hline\hline
\end{tabular*}
\label{tab:psii}
\end{table}
\begin{table}[b]
\caption{Scattering length $a_i$ and effective range $r_{i}$ for channel $i$. [in units of fm]}
\centering
\begin{tabular*}{0.5\textwidth}{@{\extracolsep{\fill}}cccc }
\toprule
$a_{1}$       & $a_{2}$~~~~~   & $r_1$~~~~~     & $r_2$~~~~~  \\
\hline
~~~$0.71$~~   & $0.50-0.16i$~~~~~   & $-0.61$~~~~~     &  $1.22-1.77i$~~~~~   \\
\hline\hline
\end{tabular*}
\label{tab:ai}
\end{table}
As shown, the probability obtained for the sum of the two $B^+D^0, B^0D^+$ channels is of the order of $96\%$.
The small deviation from unity is due to the energy dependence of the original potential of Eqs.~\eqref{eq:Vij} and \eqref{eq:pij}. 
We also observe that, while $a_1, r_{0,1}$ are real, $a_2, r_{0, 2}$ are complex because the $B^+D^0$ channel is open at the threshold of the $B^0 D^+$ channel.
The couplings are very similar, and so are the wave functions at the origin, indicating an $I=0$ state, according to Eq.~\eqref{eq:BD1}.
The results for the correlation functions of the two channels are shown in  Fig.~\ref{Fig:CF}, calculated with $R=1\, \rm fm$.

\begin{figure}[t]
  \begin{center}
  \includegraphics[scale=0.64]{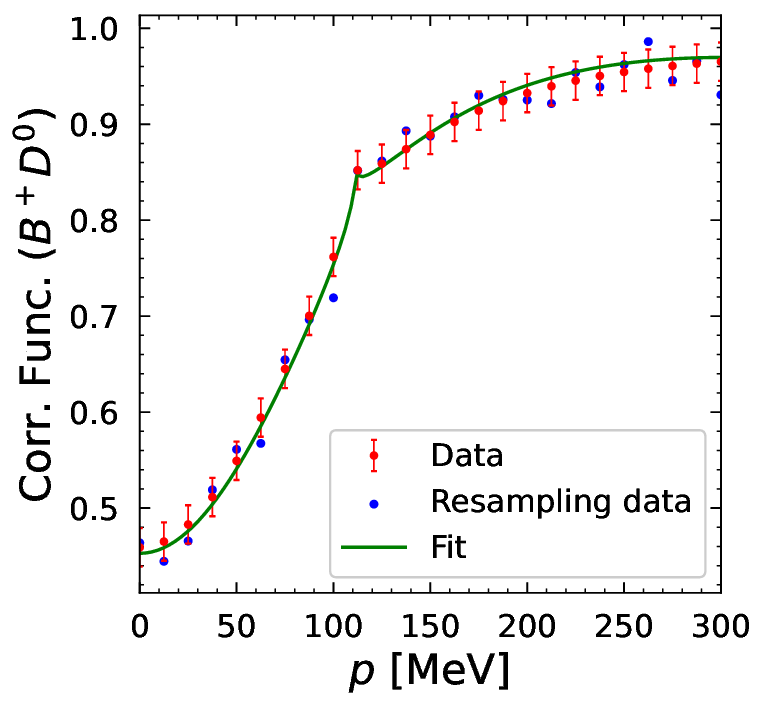}~~~
  \includegraphics[scale=0.64]{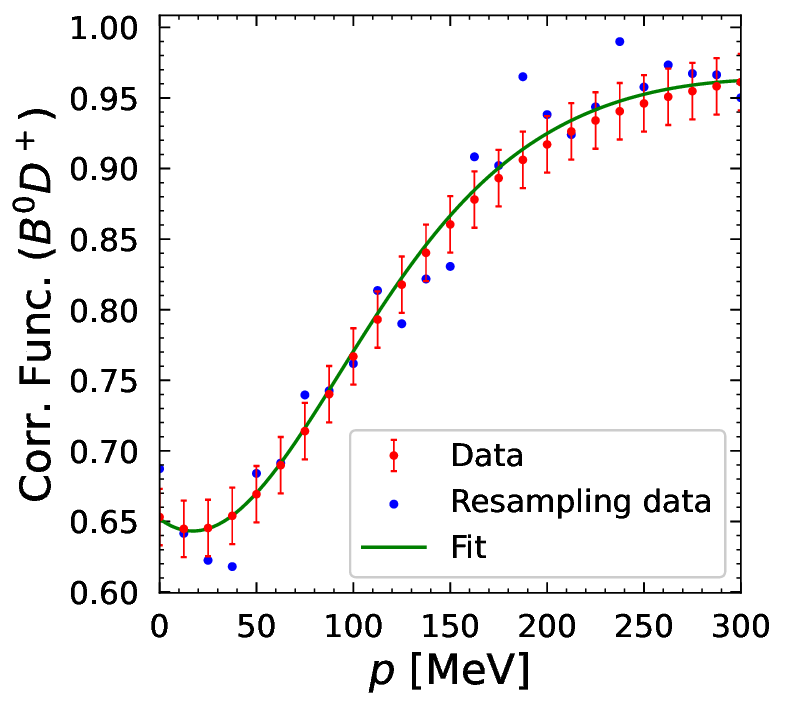}
  \end{center}
  \vspace{-0.8cm}
  \caption{Correlation functions for $B^+D^0$ (left) and $B^0D^+$ (right) channels, with 26 points in each curve with error $\pm 0.02$. The centroids of the red data follow the theoretical curve. In blue we plot the centroids obtained in one of the resampling runs, with a random Gaussian generation of the centroids of each point.}
  \label{Fig:fig2}
  \end{figure}

Next we discuss the results obtained from the resampling method fits to the data.
The data with the assumed errors are shown in Fig.~\ref{Fig:fig2}.
A warning should be given: as discussed in detail in Ref.~\cite{Feijootbb}, the sign of $V_{12}$ is undefined in the present procedure; however we rely on arguments of heavy quark flavor symmetry to choose solutions with $V_{11}$ and $V_{12}$ of the same sign.

In Table \ref{tab:Vijfit}, 
\begin{table}[t]
	\caption{Values obtained for the parameters $V'_{11}, V'_{12}$, $\alpha, \beta$, $q_{\rm max}$ and $R$.} 
\centering
\begin{tabular*}{1.0\textwidth}{@{\extracolsep{\fill}}cccccc}
\toprule
$V'_{11}$  & $V'_{12}$  &  $\alpha$  &  $\beta$  &  $q_{\rm max} \;(\rm MeV)$  &  $R \;(\rm fm)$  \\
\hline
$-1537.54\pm 918.93$   & $-1512.32\pm 913.55$      & $-31.89\pm 10.00$   &  $-165.16\pm 133.31$ &  $407.9\pm 60.8$ & $1.01 \pm 0.04$\\[2mm]
\hline\hline
\end{tabular*}
\label{tab:Vijfit}
\end{table}
we show the average values and dispersion of the parameters obtained. As we discussed above, they are not too significative due to the existing correlations.
One indication of these correlations is the relatively large parameter errors. 
Nevertheless, the important aspect is the value of the observables.
These can be found in Tables \ref{tab:gi5}, \ref{tab:psii2}, and \ref{tab:ai2}.

In Table \ref{tab:gi5},
\begin{table}[t]
  \renewcommand{\arraystretch}{0.9}
  \setlength{\tabcolsep}{0.3cm}
\centering
\caption{ Average values and dispersion of the pole position and couplings. [in units of MeV]}
\label{tab:gi5}
\begin{tabular}{ccc}
\hline
\hline
$\sqrt{s_p}$ & $g_1$  & $g_2$  \\
\hline
$7107.84\pm 17.79$  & $34623.08\pm 14300.15 $ & $34506.64 \pm 14304.57 $   \\
\hline\hline
\end{tabular}
\end{table}
we show the value of the energy at which the bound state is found, together with the values of the couplings.
As shown, a bound state is found around $7108\mev$, compatible with the bound state obtained with the original model within uncertainties.
Interestingly, the error obtained is of the order of $18\mev$.
This is not a small error for a binding energy of $39\mev$; however, this is what can be achieved with the assumed precision of the correlation data.
More  positively, using the data of the correlations at the $BD$ threshold, we are still able to predict that there is a bound state with an approximately $40\mev$ binding.

The couplings $g_1, g_2$ obtained are also compatible with the original ones and the errors are also not small.
Yet, the approximately equal values of the couplings suggest that we are dealing with an $I=0$ state.

It is interesting to analyze our obtained probabilities of the states, which are shown in Table \ref{tab:psii2}.
\begin{table}[t]
  \caption{Average value and dispersion for the probability $\mathcal{P}_i$ and wave function at the origin $\psi_i(r=0)$ for channel $i$.}
  \centering
  \begin{tabular*}{0.8\textwidth}{@{\extracolsep{\fill}}cccc }
  \toprule
  $\mathcal{P}_1$       & ~~$\mathcal{P}_2$~~   & ~~$\psi_1(r=0)$~~     & ~~$\psi_2(r=0)$~~  \\
  \hline
  ~~~$0.49 \pm 0.03$~~   & $0.42 \pm 0.03$   & $-13.61\pm 2.65$     &  $-12.61\pm 2.423$    \\
  \hline\hline
  \end{tabular*}
  \label{tab:psii2}
  \end{table}
We again obtain numbers for the probabilities of the two channels compatible with those obtained from the original model, but once again, what matters is the precision by which they can be obtained.
We observe that the uncertainties are very small, of the order of $6\%$.
This may be surprising in view of the formula used to obtain $\mathcal{P}_i$ (Eq.~\eqref{eq:Pi}), which is proportional to $g_i^2$, and $g_i$ has large errors according to Table \ref{tab:gi5}.
If $g_i^2$ is bigger in a fit because the state is more bound, $\frac{\partial G_{i}}{\partial s}$ also decreases in strength and the product becomes more stable.
This is an interesting and fortunate result of our analysis, which allows us to conclude that the application of the inverse method from the femtoscopic correlation functions would allow us to determine the nature of the bound state obtained with a high accuracy.

As shown in Table \ref{tab:ai2},
\begin{table}[t]
  \caption{Average value and dispersion for scattering length $a_i$ and effective range $r_{i}$ for channel $i$. [in units of fm]}
  \centering
  \begin{tabular*}{1.0\textwidth}{@{\extracolsep{\fill}}cccc }
  \toprule
  $a_{1}$       & $a_{2}$~~~~~   & $r_1$~~~~~     & $r_2$~~~~~  \\
  \hline
  ~~~$0.72\pm 0.03$~~   & $(0.51\pm 0.02)-(0.17 \pm 0.01)i$~~~~~   & $-0.61\pm 0.19$~~~~~     &  $(1.41\pm 0.28)-(1.65\pm 0.07)i$~~~~~   \\
  \hline\hline
  \end{tabular*}
  \label{tab:ai2}
  \end{table}
it is also rewarding to find that we can determine the scattering lengths with good precision, and the effective ranges with smaller precision but significant values.

We should also stress that the inverse method allows us to obtain the size of the source function with a relative accuracy of approximately $4\%$ (see Table \ref{tab:Vijfit}).

\section{Conclusions}
We address the problem of evaluating the correlation functions for the $BD$ system  using input extracted from a successful study of the $T_{cc}(3875)$ state. 
We take two channels, $B^0 D^+$ and $B^+ D^0$, the small mass differences between which induce visible differences in the correlation functions. 
The system also develops a bound state of approximately 40 MeV.  
Once this is achieved, we address the inverse problem of obtaining the observables associated with this system starting from the correlation functions, assuming errors as in current measurements. 
Although we obtain results compatible with those obtained from the original model, an important new result is the uncertainty by which we can obtain the observables of the system. 
We obtain the size of the source with a precision of approximately $4 \%$. 
We also determine that there is a bound state using the data of the correlation functions above the threshold of the channels, although with an uncertainty of approximately $50 \%$ of the binding energy. 
However, remarkably, we can determine the molecular nature of the obtained state with a good precision of approximately $6\%$.  
The scattering lengths of the two channels are  obtained with good precision and significant values, and the effective ranges are also obtained but with smaller precision. 
All these results indicate that the measurement of these correlation functions in the future will allow us to obtain valuable information on the $BD$ interaction and the bound states associated with this interaction. 
This study could serve as motivation to carry such measurements in the future.

\section{ACKNOWLEDGEMENT}
This work is partly supported by the National Natural Science Foundation of China under Grant No. 11975083, No. 12365019 and No. 12275076, and by the Central Government Guidance Funds for Local Scientific and Technological Development, China (No. Guike ZY22096024).
This work is also supported partly by the Natural Science Foundation of Changsha under Grant No. kq2208257
and the Natural Science Foundation of Hunan province under Grant No. 2023JJ30647 and the Natural Science Foundation of Guangxi province under Grant No. 2023JJA110076 (CWX).
This work is also partly supported by the Spanish Ministerio de Economia y Competitividad (MINECO) and European FEDER
funds under Contracts No. FIS2017-84038-C2-1-P B, PID2020-112777GB-I00, and by Generalitat Valenciana under contract
PROMETEO/2020/023.
This project has received funding from the European Union Horizon 2020 research and innovation
programme under the program H2020-INFRAIA-2018-1, grant agreement No. 824093 of the STRONG-2020 project.


%

\end{document}